\documentclass[12pt]{article}
\usepackage{graphicx}
\usepackage{amsmath}
\usepackage{amsfonts}
\usepackage{color}

\begin{document} \begin{center}
{\huge Squeeze flow of potato starch gel: effect of loading history on visco-elastic properties    }\\ \vskip .5cm
Moutushi Dutta Choudhury$^{1a}$, Shantanu Das$^{2b}$ and Sujata Tarafdar$^{1*}$\\
$^1$ Condensed Matter Physics Research Centre, Physics Department, Jadavpur University, Kolkata 700032, India\\
$^2$ Reactor Control Division, Bhabha Atomic Research Center, Trombay, Mumbai 400085, India\\
$^*$Corresponding author, sujata$\_$tarafdar@hotmail.com,
Phone: +91 33 24146666(Ex. 2760), $^a$mou15july@gmail.com, $^b$shantanu@barc.gov.in 
\end{center} 
\noindent
Abstract\\ \noindent
In this work gelatinized potato starch is shown to retain the memory of past loading history. It exhibits a visco-elastic response which does not depend solely on instantaneous conditions. A simple squeeze flow experiment is performed, where loading is done in two steps with a time lag $\tau \sim$ seconds between the steps. The effect on the strain, of varying $\tau$ is reproduced by a three element visco-elastic solid model. Non-linearity is introduced through a generalized calculus approach by incorporating a non-integer order time derivative in the viscosity equation. A strain hardening proportional to the time lag between the two loading steps is also incorporated.
 This model reproduces the three salient features observed in the experiment, namely - the memory effect, slight initial oscillations in the strain as well as the long-time solid-like response. Dynamic visco-elasticity of the sample is also reported. 

\newpage

\section{Introduction}
Complex fluids such as starch gels generally behave differently from both  Newtonian fluids  and Hookean elastic solids as discussed by Evans and Haisman(1980), Che et al. (2008) and Schofield and Blair(1937). Their rheology involves visco-elastic or visco-plastic behavior. Generalized calculus is a useful technique for describing and analyzing such materials, as shown by Das(2011), Heymans and Bauwens(1994)and Dutta Choudhury et al.(2012). One typical signature of their complexity is that in a stress-strain experiment their response at a certain time instant does not depend only on the stimulus acting at that instant,   they `remember' the history of previous treatment and respond accordingly.  We describe a simple experiment which clearly demonstrates the effect of different histories of loading on a  drop of potato starch squeezed between two parallel glass plates. 

Very simple squeeze-flow experiments on different fluids, where a drop of fluid in a Hele-Shaw cell is subjected to a load,  have shown interesting results\cite{nag}. The experiment consists of placing a weight of the order of kilograms on a drop of fluid sandwiched between two transparent plates. Spreading of the fluid is measured by video-recording the area of contact between the plate and fluid. Fluids with non-Newtonian character show novel features like oscillations in the area of contact\cite{MDC2012}. In the present experiment we report the effect of step-loading on a gelatinized starch solution. We show that when two loads are placed one after the other with a certain time lag, the asymptotic area of spreading is different from what is observed when the same two loads are applied simultaneously. We show that a three-element visco-elastic model, incorporating a non-Newtonian dashpot and a strain hardening of the elastic component, is successful in reproducing the experimental observations. The strain hardening is proportional to the time lag between the two successive loading steps. Dynamic visco-elasticity measurements have been done using a standard set up for supporting characterization of the sample. All experiments show that the sample exhibits long-time solid-like behaviour.

\section{Materials and Method}
\subsection{Sample preparation}

The non-Newtonian fluid under study is an aqueous gel of potato starch (Lobachemie, Mumbai). In our experiment,
 2.5 g of potato starch is dissolved in 100 ml of distilled water and it is heated up for 10 minutes and boiled for 2 minutes, stirring continuously, allowing it to gelatinize. The solution is cooled for 1 hour and a pinch of dye is added to enhance the contrast. 
 
\subsection{Squeeze-flow set up.}
A droplet of the non-Newtonian fluid is placed on a smooth glass plate of thickness 1.3 cm and diameter 15.5 cm. The mass of the droplet varies between 0.04 gm to 0.06 gm. A similar glass plate is placed on top of the drop. The weight of the upper plate is 0.56 kg. 
The entire experiment consists of three parts:

\begin{itemize}

\item{{\bf Experiment 1: Full loading}

30 s after  placing the upper plate, a weight $W$  is placed on it and a CCD camera (WATEC-202D, Japan) below the lower plate, records the spreading of the area of contact.  The video recording is analysed using the software Image-Pro-Plus.}

\item{{\bf Experiment 2: Step-loading-A}

The procedure is repeated with a drop of equal volume, this time the loading is done in 2 steps. 30 s after  placing the upper plate, weight $W_1$ is placed on it. After a further interval of $\tau=5.3 s$ another weight $W_2$ is added to the load on the upper plate. Here the final load $W = W_1+W_2$. The strain development in this case for $t> \tau$ is compared with the case for one-time loading.}

\item{{\bf Experiment 3: Step-loading-B}

The procedure for experiment 3 is repeated exactly, only the time lag $\tau$ between placing $W_1$ and $W_2$ on the upper plate is $\tau = 12.3 s$.}

\end{itemize}

We now compare results of the above three experiments.
Taking the instant of placing the first weight as $t=0$,
the fractional change in the area of contact of fluid and glass at time $t$

\begin{equation}
\epsilon(t) = [A(t) - A(0)]/A(0)
\end{equation}

is measured using Image-Pro-Plus software. 
Here $A(t)$ is the area at time $t$ and $A(0)$ the initial area. We identify this as the  \textit{strain} in the spreading experiment. It is difficult to make the initial area $A(0)$ exactly equal in the experiments comparing different loading histories. To overcome this problem the experiment is repeated many times and only sequences  where $A(0)$ varies within 3\% are chosen for analysis.

\subsection{Measurement of complex visco-elastic moduli}
Response of the sample to a sinusoidally varying dynamic load has been done for additional characterization of the visco-elastic behaviour. Measurements have been carried out at the Pharmaceutical Engineering Department, Jadavpur University using MCR102 SN81260812.

\section{Results}
\subsection{Squeeze flow results}

\begin{figure}[hbtp]
\begin{center}
\includegraphics[width=12.0 cm, angle=0]{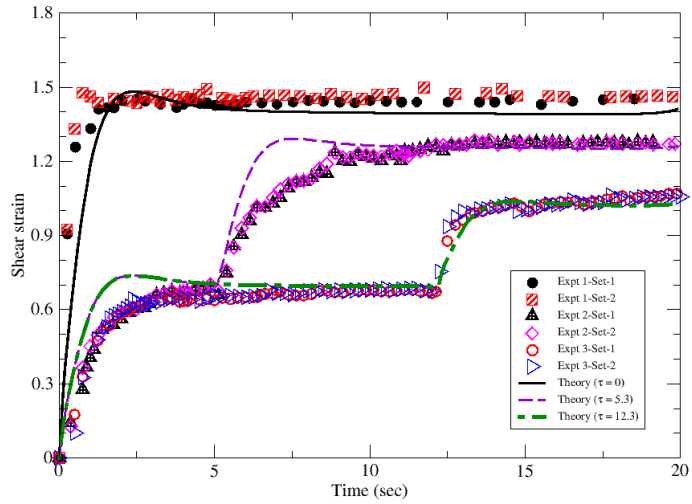}
\end{center}
\caption{Results for the strain variation with time in single step and two-step loading. Two data sets for each of the experiments are shown - solid circle and box are for single loading with 4kg, triangle up and diamond are for step loading with $\tau$ = 5.3 s; triangle right and open circle are for step loading with $\tau$ = 12.3 s. The solid line represents the calculated curves for 4 kg (single step loading), dashed line and broken line are for two-step loading with $\tau$ = 5.3 s and 12.3 s respectively, using the 3-parameter solid model.} \label{expt_theo}
\end{figure}

The strain $\epsilon(t)$  is plotted as a function of time for all 3 experiments.
The results are summarized in figure(\ref{expt_theo}). Two sets of data for $\epsilon(t)$ versus $t$ for each of experiments (1-3)  are plotted to show the reproducibility of the results. The two sets for identical conditions are designated as data sets $a$ and $b$.

In Experiment 1, $W=4.0$ kg and in Experiments 2 and 3, $W_1 = W_2 = 2.0$ kg. The time lag $\tau$ between placing the two weights is 5.3 s in Experiment 2 and 12.3 s in Experiment 3. Figure(\ref{expt_theo}) clearly shows the deviation from both ideal Hookean elastic behavior as well as from the behaviour of an ideal  Newtonian fluid. The strain does not reach the final value instantaneously as expected for a perfectly elastic material, neither does it rise from zero and continue to creep indefinitely like a perfect fluid. Moreover there is  clear evidence of history dependence in strain for $\epsilon(t)$ when $t>\tau$. The final asymptotic strain is different in all three cases - being maximum when the loading is done at once and decreasing as $\tau$ increases. Clearly, the continued loading changes the elastic and/or viscous properties of the material. We  analyse the strain build-up using different visco-elastic models and see which model is best suited to reproduce experimental results. We try to understand the characteristics of the material through these results.

\subsection{Results for dynamical loading: the complex modulus}

\begin{figure}[hbtp]
\begin{center}
\includegraphics[width=12.0 cm, angle=0]{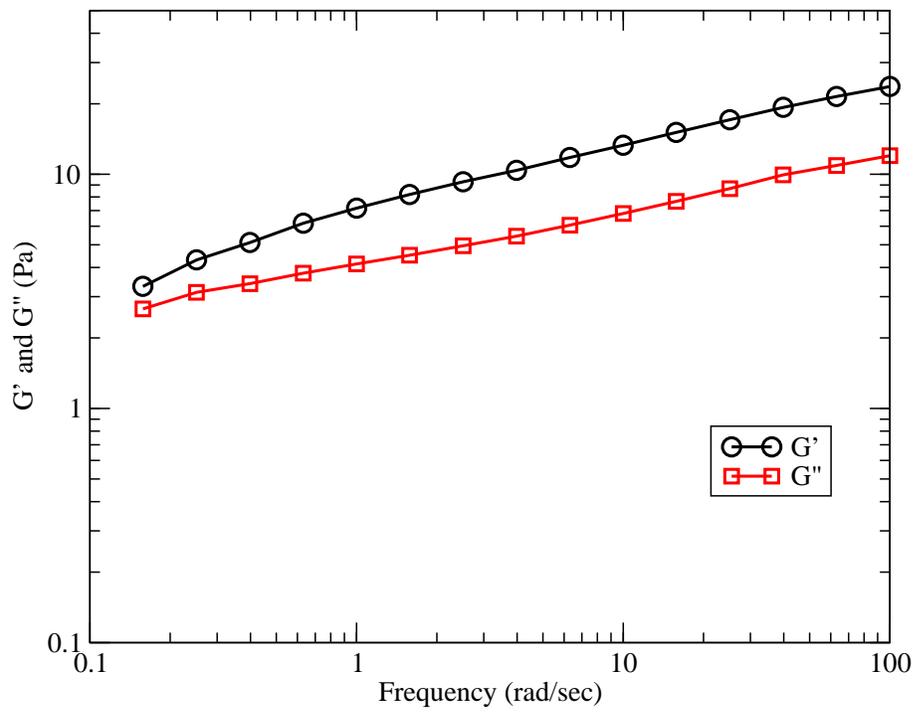}
\end{center}
\caption{Experimentally measured storage and loss modulus. Circles with solid line(black) represent $G^\prime$ and squares with red solid line represent $G^{\prime\prime}$.} \label{expt_modulus}
\end{figure}

The complex visco-elastic  modulus is written as

\begin{equation}
G^\star(\omega) = G^\prime(\omega) +iG^{\prime\prime}(\omega)
\end{equation}
In the dynamic loading experiment the sample is subjected to a sinusoidally varying strain with frequency $\omega$ rad/s and constant amplitude. Measuring the stress response which has a phase lag with the strain, gives the complex modulus $G^\star$. The real part $G^\prime$ gives the elastic contribution while the imaginary part $G^{\prime\prime}$ represents the loss modulus, i.e. the viscous part.
Graphs for $G^\prime$ and $G^{\prime\prime}$ in the frequency range 0.1 to 100 Hz are shown in figure(\ref{expt_modulus}). $G^\prime$ and $G^{\prime\prime}$ both increase with frequency and the curve for $G^\prime$ always lies above  $G^{\prime\prime}$, showing a predominantly solid-like response. This type of behaviour is typical of polymer gels\cite{wineman, melick} and starches\cite{navas}.

\section{Visco-elastic modelling}
Modelling visco-elastic materials is usually done using  energy conserving elements -  like a \textit{spring} and dissipative elements like a \textit{dash-pot}. Different combinations of these basic elements lead to different constitutive stress-strain relations. 

The simplest visco-elastic models are the Maxwell model, where a spring is in series
with a dash-pot and the Kelvin-Voigt (KV) model, where a spring and a dash-pot are
connected in parallel \cite{flugge,Bland}. These two models provide the simplest representation of a visco-elastic material. The Maxwell model behaves like a liquid on long time scales, while the KV model represents a material behaving like a solid over long times.

An infinite number of elements connected in hierarchical patterns lead to power-law relations - the hallmark of fractal systems as discussed earlier by Heymans(2003). Fractional or generalized calculus is an alternative elegant method of tackling such systems \cite{das, mainardi}.

We try to construct an appropriate model keeping in mind the salient points of the results of our experiments.
The results show the following interesting features (a) there is a finite difference between the long-time strain for the single-step loading and two-step loading experiments. The difference   increases with the time lag $\tau$ between the two loading steps, (b) immediately after loading, the initial strain shows slight oscillations as reported earlier \cite{MDC2012},(c) the strain tends asymptotically towards a more or less constant value in all three experiments, showing a solid-like nature of the material at long times.

\begin{figure}[hbtp]
\begin{center}
\includegraphics[width=12.0 cm, angle=0]{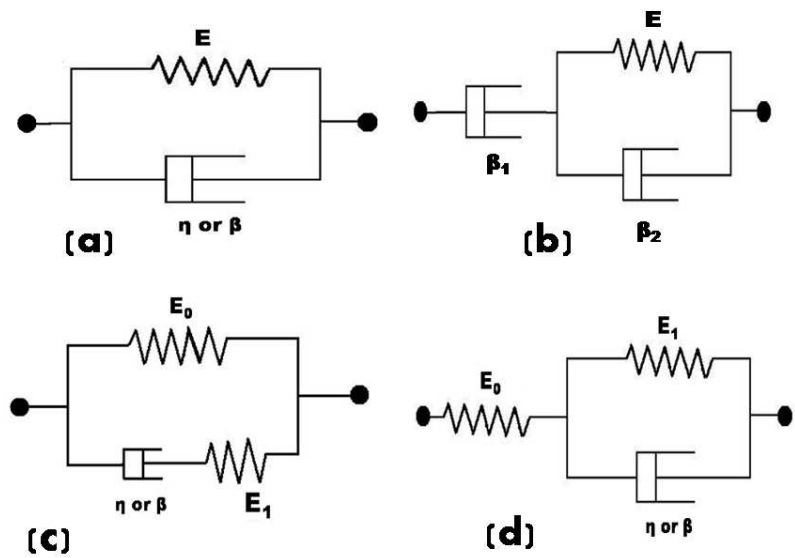}
\end{center}
\caption{Schematic diagrams of (a) the generalized Kelvin-Voigt model, (b) generalized three parameter fluid model, (c) generalized three parameter solid (Zener) model, (d) generalized three parameter solid model(Boltzman model)} \label{model}
\end{figure}

Several visco-elastic models are tried out to see whether these observations can be reproduced qualitatively. A 3-element solid-like model (the Boltzman model\cite{Chen}) is found to give the most satisfactory fit to the experimental data provided certain features are incorporated into it. These are the following - a non-Newtonian nature is attributed to the dash-pot  fluid through a non-integral time derivative in the viscosity equation; a change in the effective elastic properties of the material, similar to work hardening is introduced into the model. We allow the change in elastic modulus to increase with the time lag before the next loading step.

we describe below three different models and compare their results with our experiments.
The models are - (i) the generalized Kelvin-Voigt model, (ii) a three parameter fluid model and (iii) a three parameter solid model(Boltzman model). Fig \ref{model} shows schematic diagrams of the models (i), (ii) and two equivalent versions of model (iii).

\subsection{ (i) The generalized Kelvin-Voigt model}
\label{nonlin}

It has been shown previously \cite{MDC2012} that a Kelvin-Voigt element with a non-Newtonian viscous contribution  as shown in figure(\ref{model}$a$) can reproduce the time development of strain shown by Experiment 1, with an initially oscillating strain which saturates asymptotically. Hence we begin our analysis of the present data using this model.

In this case, the total stress $\sigma$ given below, is the sum of two terms, the first for a non-Newtonian dash-pot and the second for the spring.
\begin{equation}
 \sigma = \beta {t_c}^q \frac{d^q\epsilon}{dt^q} + E \epsilon
 \label{basic-nn}
\end{equation}
Here $\epsilon$ is the strain and $E$ is the elastic modulus. A parameter $\beta$   with dimension of stress and  a characteristic time $t_c$, characterizes the effective viscosity of the non-Newtonian fluid. For the Newtonian case $q$ would be 1 and $\beta t_c = \eta$, the coefficient of viscosity.

Solving according to the procedure outlined in Dutta Choudhury et al.(2012) we arrive at the following expression for the strain.

\begin{equation}
 \epsilon(t/t_c) = \frac{\sigma_0}{E}\left[1 - {\mathbb{E}}_q\bigg(-\frac{E}{\beta} {{(t/t_c)}^q}\bigg)\right]
\label{strain}
\end{equation}

where, $\mathbb{E}(-z)$ is the one parameter Mittag-Leffler function defined by
\begin{equation}
 \mathbb{E}_1(-z) = e^{-z}$$ $$ \mathbb{E}_q(z) = \sum_{k=0}^\infty \frac{z^k}{\Gamma(qk+1)} 
\label{Mittag}
\end{equation}

Introduction of a non-integer order derivative $q$ changes the dimensions of the physical quantities in the strain equation, so it is convenient to use non-dimensional parameters. The time has been scaled by a characteristic time $t_c$ and the strain represented as a function of the scaled time $\epsilon(t/t_c)$. The other ratios occurring in equation(\ref{strain}) -  $\frac{\sigma_0}{E}$ and $\frac{E}{\beta}$ are dimensionless.

For step loading a stress $\sigma_0$ is applied by placing the weight $W_1$ at $t = 0^+$ which produces strain
$\epsilon = \sigma_0 J(t)$. 
If at $ t = t_1$ some additional load $W_2$ is applied there is a sudden change in the strain $\epsilon(t)$\cite{flugge}. The total strain may be written as

 \begin{equation}
 \epsilon(t) =\sigma_0 J(t) + \bigtriangleup \sigma J(t-t_1)
 \label{2step}
\end{equation}
where $J(t)$ is the creep compliance at time $t$.

Hence the final strain equation for two-step loading is,

\begin{equation}
 \epsilon(t/t_c) = \frac{\sigma}{E}\left[1 - {\mathbb{E}}_q\bigg(-\frac{E}{\beta} {{(t/t_c)}^q}\bigg)\right] + \frac{\sigma}{E}\left[1 - {\mathbb{E}}_q\bigg(-\frac{E}{\beta} {{((t - t_1)/t_c)}^q}\bigg)\right]
\label{strain1}
\end{equation}

\begin{figure}[hbtp]
\begin{center}
\includegraphics[width=12.0 cm, angle=0]{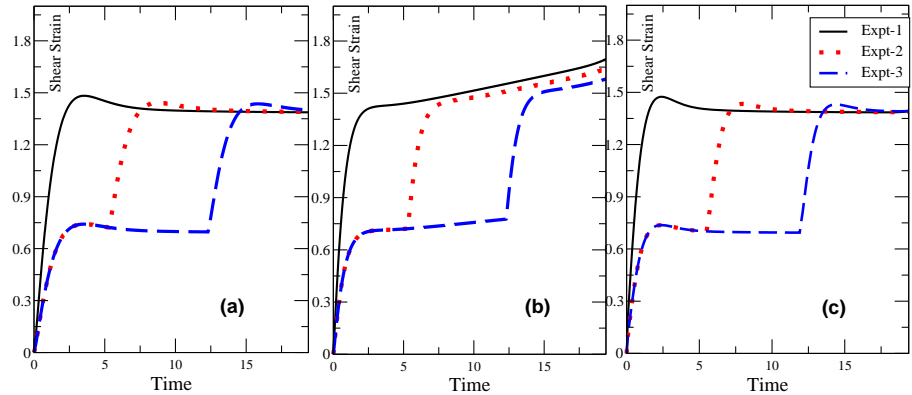}
\end{center}
\caption{The three figures from left to right, show the calculated strains for the KV model, the 3-parameter fluid model and the 3-parameter solid (Boltzman) model respectively. The black solid line represents one-step loading and dotted (red) and dashed (blue) two-step loading with $\tau=5.3 s$ and $\tau=12.3 s$ respectively.} \label{load_theo}
\end{figure}

It is to be noted here that the load on the sample due to the weight $W$ acts vertically, but the sample spreads out laterally under a shearing stress, so $\sigma$ in the equations \ref{strain} and \ref{strain1} is not numerically equal to the corresponding weight $W$, but we may assume it to be proportional to $W$. The proportionality constant is absorbed in the denominator $E$ which is also proportional but not equal to the elastic modulus of the material. This argument is discussed in detail in Dutta Choudhury et al.(2012).
Fig (\ref{load_theo}a) shows results using this model with $q=1.2$.
The initial oscillations observed when the load is placed on a drop of gelatinized potato starch is a signature of the non-Newtonian viscosity  of the fluid \cite{MDC2012}.

Theoretical results for the strain calculated from the non-linear K-V model are depicted in Fig.(\ref{load_theo}a).  
A problem with numerical methods in  fractional calculus formalism using the Mittag-Leffler function is that beyond 30 time-steps, rounding off errors accumulate and large fluctuations are observed, as discussed by Dutta Choudhury et al.(2012) and Heymans and Bauwens (1994). Up to 20 time-steps results are reliable and realistic. The time-step is scaled accordingly, so that the dimensionless time remains within this range. 

\begin{figure}[hbtp]
\begin{center}
\includegraphics[width=12.0 cm, angle=0]{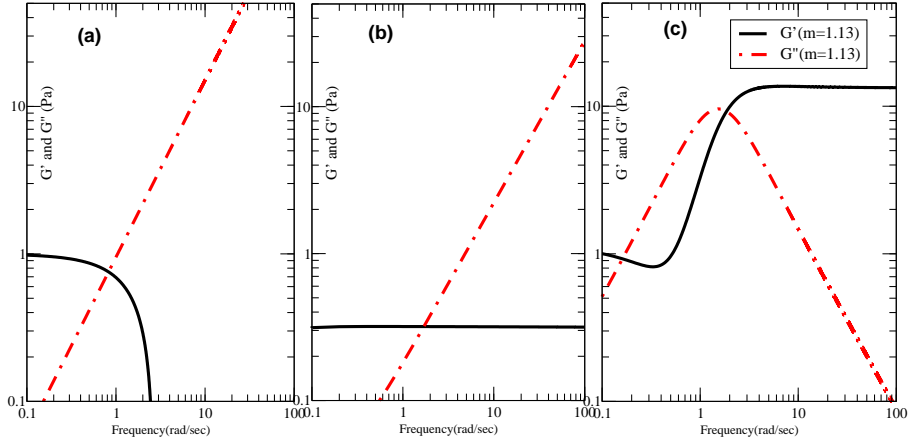}
\end{center}
\caption{The three figures from left to right, show the calculated results for the storage and loss moduli for the KV model, the 3-parameter fluid model and the 3-parameter solid (Boltzman) model respectively. The black solid line represents $G^\prime$ and the red dash-dot line represents $G^{\prime\prime}$. } \label{theo_modulus}
\end{figure}

A significant feature of our observed squeeze flow results is the `memory' retained by the material of its loading history. We can see that the KV model does not reproduce this aspect. It cannot reproduce the dynamic visco-elasticity measurements either  as evident from fig. (\ref{theo_modulus}a), so obviously a more complex model is needed. We try a 3-parameter liquid model, which can lead naturally to a difference between one time loading and step loading in the long-time limit.

\subsection{ (ii) Three parameter fluid model }

When a dash-pot is attached to the Kelvin-Voigt model in series, the resulting system represents a three-element fluid model and hence allows creep\cite{flugge}.
A linear version of the dashpots will not lead to the observed oscillations in the strain.
To introduce non-linearity in the viscous part of 3 parameter fluid model, we replace the linear dash-pots by fractional order dash-pots, the outer dash-pot incorporates a fluid  with the fractional order $q_1$ for the time derivative and the dash-pot in the K-V model is substituted by a $q_2$ order dash-pot. So the fractional order constitutive equation reduces to,

\begin{equation}
\bigg[ 1 + (\beta_1/E) \frac{d^{q_1}}{dt^{q_1}} + (\beta_2/E) \frac{d^{q_2}}{dt^{q_2}}\bigg] \sigma(t/t_c) = \bigg[ b_1 \frac{d^{q_1}}{dt^{q_1}} + (\beta_1 \beta_2/E) \frac{d^{q_1+q_2}}{dt^{q_1+q_2}}\bigg] \epsilon(t/t_c)
\label{NL3parafluid}
\end{equation}

where,$\beta_1$ and $\beta_2$ are the apparent viscosities of outer dash-pot and the dash-pot in K-V model of figure (\ref{model}b) and $E$ is the elastic modulus of the spring part of the model.

Taking it's Laplace transform and solving we get the strain-time relation as
\begin{equation}
\epsilon(t/t_c) = \sigma_0 / \beta_1 \frac{t^q_1}{\gamma(q_1 + 1)} + \sigma_0 / \beta_2  \mathbb{E}_2(q_2,\alpha)
\label{solNL3parafluid}
\end{equation}

where $\mathbb{E}_2(q_2,\alpha)$ is the second order Mittag-Leffler function defined as ,

\begin{equation}
\mathbb{E}_2(q_2,\alpha) = \sum_{k=0}^\infty \frac{(-\alpha)^k{(t/t_c)^{(k + 1)q_2}}}{\Gamma[(k+1)q_2 + 1]} 
\label{Mittag1}
\end{equation}

The equation for the final strain after total loading, 

\begin{equation}
\epsilon(t/t_c) = \sigma_0 / \beta_1 \frac{t^q_1}{\gamma(q_1 + 1)} + \sigma_0 / \beta_2  \mathbb{E}_2(q_2,\alpha) + \sigma_0 / \beta_1 \frac{(t - t_1)^q}{\gamma(q_1 + 1)} + \sigma_0 / \beta_2  \mathbb{\textit{E1}}_2(q_2,\alpha)
\label{sol3}
\end{equation}

Here 

\begin{equation}
\mathbb{\textit{E1}}_2(q_2,\alpha) = \sum_{k=0}^\infty \frac{(-\alpha)^k{(t-t_1)/t_c^{(k + 1)q_2}}}{\Gamma[(k+1)q_2 + 1]} 
\end{equation}

i.e, $t$ in (\ref{Mittag1}) is replaced by $(t-t_1)$.

For step loading we again follow the treatment described earlier to get the plot of $\epsilon(t/t_c)$ against $t$ as shown in figure(\ref{load_theo}b). The graph shows that the system remembers the loading history of the material i.e. in the asymptotic long-time limit there remains a gap between the strain curves for one-time loading and step-loading.
However, the creep introduced by this model is not observed in the experiments, rather the strain saturates to a constant value for both one-time loading and step-loading.
As these results do not fit our experiments satisfactorily, we revert to a solid-like model with some modifications as described below.

\subsection{(iii) Three parameter solid (Boltzman) model with non-linear viscous elements}
\label{3para_nonlin}

The third model we consider is the Boltzman model \cite{Chen} a 3-parameter visco-elastic model where the Kelvin-Voigt solid and a spring are connected in series as shown in fig.(\ref{model}d).This model has been shown to be equivalent to the Zener  model [fig.(\ref{model}c)]. Such models have been used to describe visco-elastic behaviour of biological materials such as \textit{Cartilage} and \textit{white blood cell membrane} \cite{Chen}.

\begin{table}
\label{tab_1}
\begin{tabular}{|c|c|c|c|c|}
\hline Model & For Full  & 1st step-loading & 2nd step-loading & 2nd step-loading \\ 
 parameter & loading & $\tau$=0s & at $\tau$=5.3s (A) & at $\tau$=12.3s (B) \\ 
\hline $\sigma$ & 1.56 & 0.78 & 0.78 & 0.78 \\ 
\hline $q$ & 1.2 & 1.2 & 1.2 & 1.2 \\ 
\hline $a_1$ & 0.6 & 0.6 & 0.6 & 0.6 \\ 
\hline $b_1$ & 8 & 8 & 8 & 8 \\ 
\hline $m$ & 1.13 & 1.13 & 1.30 & 2.78 \\ 
\hline $\alpha$ & 1.6 & 1.6 & 1.6 & 1.6 \\ 
\hline 
\end{tabular} \caption{The  parameters for the 3-parameter Boltzman model under different loading sequences.}
\end{table}

The Boltzman model shows visco-elasticity with long-time solid like behaviour. This model can be shown to be equivalent to the Zener model, which also consists of a dashpot and two springs ( see fig.(\ref{model}(c and d))).
Our trials with the previous models lead us to the conclusion that we cannot model the memory of loading history using a solid-like model, assuming a constant set of model parameters. The phenomenon of strain-hardening or work-hardening, where visco-elastic properties change on loading is well known \cite{degarmo, schofield}. We now introduce this aspect into our visco-elastic model. The primary idea is that during the second step for two-step loading, the material has already been under stress for time $\tau$ and this causes a change in its elastic modulus, the change being proportional to the time interval. So, while the one-time loading and first step of the two-step loading are under identical conditions, a change in material properties has to be considered during the second loading step.
The viscous element - the dash-pot is replaced by its intermediate Scott-Blair fractional system of order q \cite{das}. This concept is similar to the section\ref{nonlin}.
The constitutive equation for the fractional Boltzman model is,

\begin{equation} 
 [1 + a_1 \frac{d^q}{dt^q}] \sigma(t/t_c) = [m + b_1 \frac{d^q}{dt^q}] \epsilon(t/t_c)
\label{NLS}
\end{equation}

with 
$a_1 = \frac{\beta}{E_0 + E_1}$, 
$m = \frac{E_0 E_1}{E_0 + E_1}$ and
$b_1 = \frac{E_0 \beta}{E_0 + E_1}$

the various elements are shown in the figures (\ref{model}c) and (\ref{model}d).
A parameter $\beta$ characterizes the effective viscosity of the non-Newtonian fluid and when $q$ is 1, $\beta t_c = \eta$, the coefficient of viscosity.
Proceeding as in section (\ref{nonlin}), we get the solution of (\ref{NLS}),
  
 \begin{equation}
 \epsilon(t/t_c) = \sigma_0 \frac{a_1}{b_1}+\sigma_0\bigg(\frac{1}{m}-\frac{a_1}{b_1}\bigg)[1 - {\mathbb{E}}_q\bigg(-{(t\alpha)}^q\bigg)]
  \label{solNLS}
\end{equation}
 
 The material functions are,
 
 \begin{equation}
 J(t/t_c) = J_g + J_1\bigg[1 -{\mathbb{E}}_q\bigg(-(t\alpha)^q\bigg)\bigg]
 \label{solNLS1}
\end{equation}

$J_g = \frac{a_1}{b_1}$; $J_1 = [\frac{1}{m} - \frac{a_1}{b_1}]$ ; and $\alpha = \frac{m}{b_1}$ contains the relaxation time $t_c$ which is taken as 1 here.
 For getting the total strain after step-loading we follow equation (\ref{2step}) and finally reach the solution as,
 \begin{equation}
 \epsilon(t/t_c) =\sigma_0 J(t/t_c) + \bigtriangleup \sigma J[(t-t_1)/t_c]
\end{equation}

We assume that \textit{work hardening} or  \textit{lethargy} develops in the material in the second step during step-loading. This is taken into account by assigning a stiffer elasticity to the sample through a modified value of $m$, which contains the combination of $E_0$ and $E_1$. 
The best-fit parameters for model (iii) are given in table(\ref{tab_1}) and 
 calculated curves are compared with experimental results in fig.(\ref{expt_theo}).

\subsection{Calculation of dynamic moduli $G^{\prime}(\omega)$ and $G^{\prime\prime}(\omega)$}

Expressions for $G^{*} = G^{\prime}(\omega) + i G^{\prime\prime}(\omega)$ for the three models described above are as follows:

\subsubsection{Fractional Kelvin-Voigt model}

We have calculated complex relaxation modulus from the Kelvin-Voigt fractional-model. The form of this modulus is:
\begin{equation}
G^{*} = (E + \beta \omega^{q} \cos{\frac{q\pi}{2}}) + i(\beta\omega^{q} sin{\frac{q\pi}{2}})
\end{equation}

Here $G^{\prime} = E + \beta \omega^{q} \cos{\frac{q\pi}{2}}$ and $G^{\prime\prime} = \beta\omega^{q} sin{\frac{q\pi}{2}}$ are the elastic modulus and loss modulus respectively.

\subsubsection{3-parameter non-linear fluid model}
The second model i.e. the 3-parameter non-linear fluid model reveals complex features of elastic modulus and loss modulus as
\begin{equation}
G^{\prime} = \frac{B \cos{\frac{p\pi}{2}} (\alpha^{2} + \omega^{n+q} + 2\alpha\omega^{n}\cos{\frac{q\pi}{2})} + C \omega^{2p} (\alpha + \omega^{q}\cos{\frac{p\pi}{2})}}{D}
\end{equation}

and

\begin{equation}
G^{\prime\prime} = \frac{B \omega^{q}(\omega^{2q}\sin{\frac{q\pi}{2}} + \alpha^{2} \sin{\frac{p\pi}{2}} + 2\alpha\omega^{q}\sin{\frac{p\pi}{2}}\cos{\frac{q\pi}{2}}) + C \omega^{n+p}\sin{\frac{q\pi}{2}}}{D}
\end{equation}

where $n = p+q$, 
$ B = \beta_{1}\beta_{2}^{2}$, 
 $C = \beta_{1}^{2}\beta_{2}$
 and $D = \beta_{1}^{2}\omega^{2p} + \beta_{2}^{2}\omega^{q}(\omega^{q} + \alpha^{2} + 2\alpha\cos\frac{q\pi}{2}) + 2\beta_{1}\beta_{2}(\omega^{n}\cos{\frac{(p-q)\pi}{2}} + \alpha\omega^{p}\cos{\frac{p\pi}{2}})$
 
\subsubsection{3 parameter Boltzman solid model}
Complex relaxation modulus after taking the Laplace transform of constitutive equation(\ref{NLS}) for 3 parameter Boltzman model,

$G^*(\omega) = \phi(s)\mid_{s=i\omega}=\frac{\sigma(s)}{\epsilon(s)}$\\
gives\\
$G^*(\omega)= G^{\prime}(\omega) + i G^{\prime\prime}(\omega)$\\
where
\begin{equation}
G^\prime(\omega) = \frac{m +(a_1 m + b_1)\omega^q \cos q\pi/2 + a_1 b_1 \omega^{2q}}{1 + 2 a_1 \omega^q \cos{q\pi/2} + a_1^2 \omega^{2q}}
\end{equation}

and
\begin{equation}
G^{\prime\prime}(\omega) = \frac{(b_1 - m a_1)\omega^q \sin{q\pi/2}}{1 + 2 a_1 \omega^q \cos{q\pi/2} + a_1^2 \omega^{2q}} 
\end{equation}

$G^\prime$ and $G^{\prime\prime}$ are plotted in Fig.(\ref{expt_modulus}) for the three models. $G^{\prime\prime}$ increases linearly in the first two models (a) and (b), and shows a peak in the 3-parameter solid model. $G^\prime$ is a non-increasing function of frequency in the first two models, but exhibits a complex variation in the third model with a generally increasing trend.

\section{Discussion: comparing theory and experiment}
Let us compare the performance of the three different models presented in reproducing experimental results.

The extended KV model [\ref{model}a], with non-linear viscosity, successfully explained the oscillating strain for single step loading in the earlier work \cite{MDC2012}. However, being a solid like model it cannot lead naturally to a long-time difference between single-step and multi-step loading, under identical total load \cite{flugge}. The dynamic viscosity trends are also totally different from the experimentally obtained curves. So obviously a more complex model is needed to reproduce the new results.

The three-parameter fluid model (\ref{model}b) can reproduce naturally the asymptotic difference between single-step and multi-step loading. But, being essentially a liquid-like model, it shows creep. The strain continues to increase approximately linearly at long times, this is not in conformity with experiments. The experimental storage and loss moduli also show a
dominance of the elastic i.e. solid-like behaviour compared to the dissipative, i.e. loss modulus. So, even with three elements and non-linear viscosity the three-parameter fluid model fails to reproduce the features seen in the real system.

It is now obvious that to simultaneously exhibit a memory of loading history and long-term saturation of strain, a solid-like model with additional modification is necessary. A lethargy effect or strain hardening has been observed in polymeric as well as crystalline materials \cite{degarmo} and is quite likely to occur in a complex material like a starch gel.  The non-linear viscosity in the model leads to the oscillatory variation seen after initial application of the load.
 Introducing these in the three-parameter Boltzman solid model (d) we get  strain variation curves (represented by lines) which can reproduce the experimental results (denoted by the symbols) quite well as shown  in fig (\ref{expt_theo}).
The dynamic response of the system under oscillatory strain has been calculated for the three models and compared with experiments. Figures (\ref{theo_modulus}) show that experimental variations of storage and loss moduli do not match any of the calculated results exactly. However, models (a) and (b) yield very simple curves for $G^\prime$ and $G^{\prime\prime}$ with $\omega$ where there is no indication of any increasing trend in $G^\prime(omega)$. Model (d) i.e. the modified three-parameter  solid model shows a general rising trend for $G^\prime(omega)$, particularly noticeable in a certain frequency range. Though this model does not lead to the more-or-less linear rise seen in the experiment, it is closer than the other models. $G^{\prime\prime}$ does not exhibit a peak in the experiment, as seen in model (d) results, but it is possible that the frequency range accessible in the experiment is inadequate for this. Evidently, more modifications of the model are needed, for faithful quantitative representation of dynamic visco-elasticity behaviour. For the present we may consider model (d) a good starting point. Other reports on potato starch and other starches show a behaviour similar to our experimental results \cite{navas}.

\section{Conclusions}

The interesting visco-elastic behaviour of starch gels, particularly potato starch is manifested in several very simply performed experiments. Micrographs show potato starch to consist of rounded cellular structures \cite{leach, Eliasson} which are known to contain amylose and amylo-pectin. On heating, the long polymer chains of amylose leach out of the cells,causing a thickening termed gelatinization. The highly viscous fluid produced exhibits non-linear viscoelasticity and other forms of instability. For example, finger-like undulations develop at the  periphery of a squeezed drop(\cite{MDC2}) and there are oscillations in the area of contact of the drop and substrate. Potato starch gel also acts as a host for crystal growth and a drying drop of the gel with added NaCl can produce salt crystals with a wide range of morphologies \cite{mou}.

These results are important in food industry \cite{navas}, related biological systems\cite{meral} as well as from the general point of view of visco-elasticity. Considering the host of interesting phenomena observed in such very simple experiments, studies of the type  reported here promise to be very useful.

\section{Acknowledgment}
MDC is grateful to CSIR, India for award of a Senior Research Fellowship.

\end{document}